\documentclass[%
 reprint,
 amsmath,amssymb,
 aps,
]{revtex4-1}
\usepackage{graphicx}
\usepackage{color}
\usepackage{placeins}
\usepackage{float}
\usepackage{tabularx,colortbl}
\usepackage{psfrag}
\usepackage{subfigure}
\usepackage{amsmath}
\usepackage{setspace}
\usepackage{graphicx}
\usepackage{dcolumn}
\usepackage{bm}

\begin{document}

\preprint{APS/123-QED}

\title{Generalized Spatial Talbot Effect based on All-dielectric Metasurfaces}

\author{Shulabh Gupta}
\email{shulabh.gupta@carleton.ca}
\affiliation{%
 Department of Electronics, Carleton University\\
 1125, Colonel by Drive, Ottawa, Ontario, Canada.
}%


\begin{abstract}
A generalized spatial Talbot effect is proposed where the period of the input aperture is scaled by a non-integer real value, as opposed to the integer-only factor in a conventional Talbot effect. This is achieved by allowing and engineering the phase discontinuity distributions in space using metasurfaces, in conjunction with free-space propagation. The introduction of such abrupt phase discontinuities thereby enables non-integer scalings of the aperture periodicities. Specific implementations using Huygens' metasurfaces are proposed and their operation to achieve non-integer scaling of the input aperture period is demonstrated using numerical results based on Fourier propagation. 
\end{abstract}

\pacs{Valid PACS appear here}
\maketitle


\section{Introduction}

Talbot effect was originally observed by Henry Fox Talbot and described in his seminal work in \cite{Talbot_Original}. When a plane wave is incident on a periodic aperture, the image of the aperture is self-replicated at specific discrete locations away from the aperture. At other distances in between, the aperture is self-imaged with smaller period resulting in higher repetition of the periodic illuminations. This effect is known as the spatial Talbot effect. Talbot effect has found extensive applications in imaging, optical communication, optical computing, and optical interconnection, to name a few and a more comprehensive review can be found in \cite{Talbot_Review}. The Talbot effect has also been translated to temporally periodic signals. In a temporal Talbot effect, the space-time duality is exploited to achieve self-imaging of periodic pulse trains, where their repetition rates are increased using simple phase-only filtering techniques \cite{TemporalTalbot}\cite{Azana_Talbot_Temporal}.

While Talbot effect has conventionally been used in repetition-rate multiplication of either the periodic spatial aperture or the periodic pulse trains, they have been extended to achieve repetition-rate division as well recently \cite{Talbot_Amplification}\cite{Talbot_Averaging_Jose}. However, to the best of my knowledge, the period of the aperture is either multiplied or divided by an \emph{integer} factor only. In this work, this concept of period scaling is  generalized to \emph{non-integer} factors, where the input periodic aperture can be imaged with either a higher or a lower repetition rate by any arbitrary real number. The specific contributions of this work are: 1) Proposal of a generalized Talbot effect described in the spatial domain for arbitrary scaling of aperture periodicities. 2) Specific implementations of generalized Talbot system using metasurfaces \cite{meta2}\cite{meta3}, which are 2D array of subwavelength particles to provide abrupt phase discontinuities in space.

\section{Principle}

\subsection{Conventional Talbot Effect}

\begin{figure}[htbp]
\begin{center}
\psfrag{c}[c][c][0.7]{$|\psi(x,y,0)|$}
\psfrag{e}[c][c][0.7]{$|\psi_T(x,y,\Delta z)|$}
\psfrag{f}[c][c][0.7]{$\psi(x,y,z)$}
\psfrag{x}[c][c][0.7]{$x$}
\psfrag{y}[c][c][0.7]{$y$}
\psfrag{z}[c][c][0.7]{$z$}
\psfrag{n}[c][c][0.7]{Periodic Aperture}
\psfrag{o}[l][c][0.7]{$\Lambda = m X_0\quad\text{where}\quad m\in\mathcal{R}$}
\psfrag{p}[l][c][0.7]{$\Lambda = X_0/m\quad\text{where}\quad m\in\mathcal{I}, \ge1$}
\psfrag{s}[c][c][0.7]{$X_0$}
\psfrag{t}[c][c][0.7]{$\Delta z$}
\psfrag{A}[c][c][0.7]{$T(x,y)$}
\includegraphics[width=0.9\columnwidth]{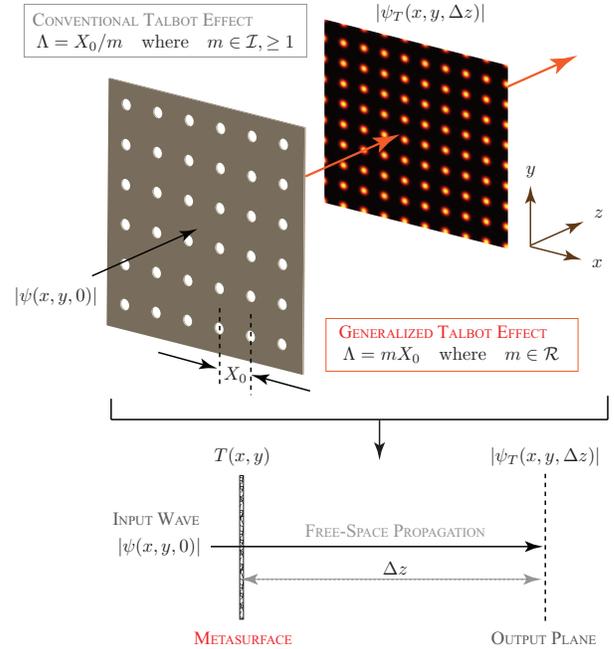}
\caption{Generalized spatial Talbot effect where the period of the 2D array of sources is scaled by an arbitrary real number $m$ in the output plane, as opposed to integer-only values in the conventional Talbot effect.}\label{Fig:Principle}
\end{center}
\end{figure}

Consider an aperture consisting of a 2D array of holes with period $\Lambda = X_0$ at $z=0$, as shown in Fig.~\ref{Fig:Principle}(a), illuminated by a plane wave of frequency $\omega$ (or wavelength $\lambda$). Such an aperture acts as a 2D array of point sources, which then radiate along the $z-$axis. Due to diffraction, the periodic sources interfere with each other forming complex diffraction patterns. However, at integer multiples of distances $\Delta z  = X_0^2/\lambda = z_T$, known as the Talbot distance, the input aperture distribution is re-constructed with the same periodicity as a result of free-space interference. This phenomenon is called the integer Talbot effect. At other distances $z = z_T/m$, the fields are self-imaged with $m-$times the spatial repetition rate of the input aperture, as shown in the bottom of Fig.~\ref{Fig:Principle}(a), i.e. $\Lambda = X_0/m$. This phenomenon is called the fractional Talbot effect. Therefore, in a conventional spatial Talbot effect, the periodicity of the input aperture is always increased by an \emph{integer} number $m$, i.e.

\begin{equation}
\overbrace{\psi_T(x,y,z_T/m)}^{\Lambda = X_0/m}= \underbrace{\psi_T(x,y,0)}_{\Lambda = X_0} \ast h(x,y),\label{Eq:ConventionalTalbotEffect}
\end{equation}

\noindent where $h(x,y)$ is the impulse response of free-space.

\subsection{Generalized Talbot Effect}

The conventional Talbot effect is observed under the assumption that there is no phase discontinuity across the aperture so that $\psi_T(x,y,0) = |\psi_T(x,y,0)|$. This assumption consequently restricts the repetition-rate increase factor $m$ to integer values only. However, if the aperture is allowed to feature abrupt phase discontinuities across it, this restriction of integer-only values of $m$ is lifted. This can be understood by considering two fractional Talbot distances $z_1 = z_T/p$ and $z_2 = z_T/q$ such that

\begin{align}
\psi_T(x,y,z_T/p) = |\psi_T(x,y,0)|\ast h(x,y)\\
\psi_T(x,y,z_T/q) = |\psi_T(x,y,0)|\ast h(x,y).
\end{align}

\noindent Alternatively, the above equations can be written in a complex spatial frequency domain as

\begin{align}
\tilde{\psi}_T(k_x,k_y,z_T/p) = |\tilde{\psi}_T(k_x,k_y,0)| \times \tilde{H}(k_x,k_y)\\
\tilde{\psi}_T(k_x,k_y,z_T/q) = |\tilde{\psi}_T(k_x,k_y,0)| \times \tilde{H}(k_x,k_y).
\end{align}

\noindent Substituting $|\tilde{\psi}_T(x,y,0)|$ from the second equation into the first and using $\tilde{H}(k_x,k_y,z ) = e^{-ik_z z}$, one get

\begin{align}
\tilde{\psi}_T(k_x,k_y,z_T/p) &= \tilde{\psi}_T(k_x,k_y,z_T/q) \exp\left[-ik_z z_T\left(\frac{1}{p}- \frac{1}{q}\right)\right]
\end{align}

\noindent In this case considered, $p < q$. Inverse Fourier transforming the above equation leads to

\begin{align}
\psi_T(x, y ,\Delta z) &= \psi_T(x,y,0) \ast h(x,y, \Delta z),
\end{align}

\noindent where $\Delta z = z_T(1/p - 1/q)$, where $z= z_T/q$ is chosen as a new reference plane. This equation can then be further written as

\begin{align}
\overbrace{\psi_T(x, y ,\Delta z)}^{\Lambda= X_0/p} &= [\underbrace{|\psi_T(x,y,0)|}_{\Lambda= X_0/q} \times \overbrace{T(x,y)}^\text{Metasurface}]  \ast h(x,y, \Delta z),
\end{align}

\noindent where $T(x,y)  = \angle \psi_T(x,y,0) = \angle \psi_T(x,y,z_T/q)$. This equation tells us that an input aperture field with a period $\Lambda= X_0/q$ when multiplied with a phase function $T(x,y)$ results in another periodic field at $z=\Delta z$ with a period $\Lambda= X_0/p$. In other words, the period has been scaled by a non-integer value $m = q/p$ between the input and the output planes. Such a system is referred here to as the \emph{generalized spatial Talbot system} and the corresponding effect as the generalized Talbot effect, as illustrated in the bottom of Fig.~\ref{Fig:Principle}(b). The phase function $T(x,y)$ represents a spatial phase discontinuity profile which enables the scaling of the repetition rate of the input aperture fields by a non-integer value. Such a phase discontinuity can be easily introduced using a metasurface, as will be shown in Sec. IV. 

In summary, when the input aperture field with the period $\Lambda = X_0$ is spatially cascaded with a metasurface with transmittance $T(x, y) = \psi_T(x,y,z_T/q)$, and propagated by a distance $z = z_T(1/p - 1/q)$, the output field distribution exhibits a scaled period $\Lambda = (q/p) X_0$.

\section{Metasurface Transmittance Functions}

\begin{figure*}[htbp]
\begin{center}
\psfrag{a}[c][c][0.7]{$y~(\mu\text{m})$}
\psfrag{b}[c][c][0.7]{$x~(\mu\text{m})$}
\psfrag{c}[c][c][0.7]{$|\psi(x,y,0)|^2$~(dB)}
\psfrag{d}[c][c][0.7]{$\angle T(x,y)$~($ \pi$~rad)}
\psfrag{e}[c][c][0.7]{$|\psi(x,y,z_1)|^2$~(dB)}
\psfrag{h}[c][c][0.7]{$|\psi(x,y,\Delta z)|^2$~(dB)}
\psfrag{f}[c][c][0.7]{$\boxed{1/3~\text{Division}}$}
\psfrag{g}[c][c][0.7]{$\boxed{1/4~\text{Division}}$}
\psfrag{n}[c][c][0.7]{\color{white}{$-4.15$~dB}}
\psfrag{m}[c][c][0.7]{\color{white}{$-3.5$~dB}}
\psfrag{p}[c][c][0.7]{$\Delta z = (1/3 - 1/4)z_T$}
\psfrag{q}[c][c][0.7]{$\Delta z = (1/2 - 1/3)z_T$}
\psfrag{r}[c][c][0.7]{$\boxed{m = 4/3 = 1.33}$}
\psfrag{s}[c][c][0.7]{$\boxed{m = 3/2 = 1.50}$}
\includegraphics[width=1.7\columnwidth]{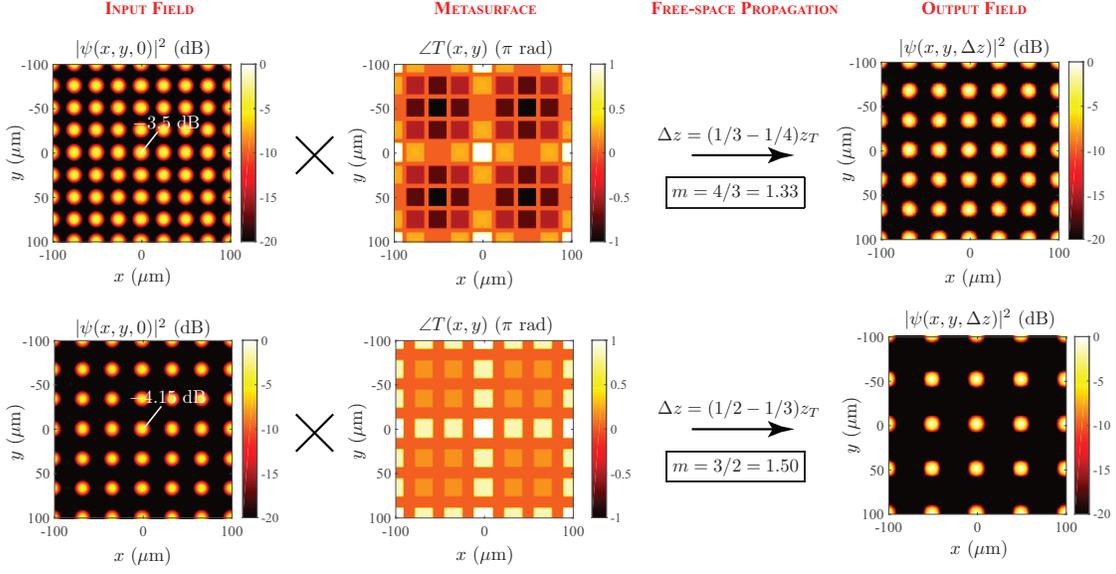}
\caption{Computed fields at the output plane when a periodic array of sources are phased following \eqref{Eq:PhaseEven} and \eqref{Eq:PhaseOdd}, to achieve a repetition rate scaling by a factor of $m=1.33$ and $m=1.5$, respectively. Here $X_0=100~\mu$m and the individual sources are assumed to be Gaussian functions $\psi(x,y) = \exp[-(x^2 + y^2)/2w_0^2]$, with $w_0 = 5~\mu$m. The design frequency is $250$~THz. Only a small part of the overall aperture is shown for clarity.}\label{Fig:IdealResults}
\end{center}
\end{figure*}

As described in the previous section, the metasurface of Fig.~\ref{Fig:Principle}, is a phase-only function, and it mimics the phase distributions of the fractional Talbot self-images, i.e. $T(x,y)  =  \exp\{i\angle \psi_T(x,y,z_T/m)\}$. In this section, the closed-form expressions of these phase distributions will be developed.

Let us assume for simplicity, a one-dimensional periodic array of sources with the amplitude distribution

\begin{align}
\psi_\text{in}(x) &= \sum_{a = -\infty}^{+\infty} \delta(x-aX_0).\label{Eq:Input}
\end{align}

\noindent After a propagation through free-space along $z-$axis, the output fields are given by $\psi(x) = \psi_\text{in}(x)\ast h(x,z)$, where the impulse response $h(x)= \exp(-i\pi x^2/\lambda z)$, under paraxial conditions (time convention used here is $e^{j\omega t}$). Using this equation with \eqref{Eq:Input}, and simplifying the convolution integral, we get

\begin{equation}
\psi(x) = e^{-i\frac{\pi x^2}{\lambda z}} \sum_{a = -\infty}^{+\infty} \exp\left(-i\pi a^2 n\right)\exp\left(in\frac{2a \pi  }{X_0}x\right),\label{Eq:TalbotProp}
\end{equation}

\noindent where $X_0^2/\lambda z = n$ assumed to be an integer, i.e. $n\in \mathcal{I}$. This corresponds to propagation distance $z= z_T/n$, where $z_T = X_0^2/\lambda$, known as the Talbot distance. 

\begin{enumerate}
\item Case 1: when is $n$ \emph{even}, $\exp\left(-i\pi a^2 n\right) = 1$, and thus \eqref{Eq:TalbotProp} reduces to 

\begin{align}
\psi(x) & = \frac{X_0}{n} \sum_{a =-\infty}^{+\infty} \delta\left(x-a\frac{X_0}{n}\right)\exp\left(-i\frac{\pi a^2}{n}\right), \label{Eq:Even}
\end{align}

\item Case 2: when $n$ is \emph{odd}, \eqref{Eq:TalbotProp} reduces after straightforward manipulation to 

\begin{align}
\psi(x) =  \frac{X_0}{n}\sum_{a= -\infty}^{+\infty} \delta\left(x-a\frac{X_0}{2n}\right) [1 -e^{j\pi a}]\exp\left(-i\frac{\pi a^2}{4n}\right) \label{Eq:Odd}
\end{align}

\end{enumerate}

These equations are derived using the identity $\sum_{a=-\infty}^{\infty} \exp(2\pi j ax/X_0) = X_0 \sum_{a=-\infty}^{\infty} \delta(x - aX_0)$. Equations \eqref{Eq:Even} and \eqref{Eq:Odd} reveals that the period of the output field $\psi(x)$ is smaller by a factor of $n$ compared to the input field in both cases, as expected at the fractional Talbot distance $z= z_T/n$. Repeating the same procedure for a 2D array of sources, the complex phase of the output self-imaged patterns at the location $(a,b)$ on the aperture can be verified to be \cite{Talbto_Phase_Analytical}

\begin{subequations}
\begin{equation}
\phi(a, b) = -\frac{2\pi (a^2 + b^2)}{n}\label{Eq:PhaseEven}
\end{equation}
\begin{equation}
\phi(a, b) = -\frac{\pi [(2a+1)^2 + (2b+1)^2]}{4n}\label{Eq:PhaseOdd}
\end{equation}
\end{subequations}

\noindent These equations can be used to determine the complex phase of the fields at any fractional Talbot distance lying between $[0,\; z_T]$, and thus can be used to construct the metasurface transmittance function $T(x,y)$ for the case of generalized Talbot effect as described in Sec. II. It should be noted that while the above phase functions are developed only for the special case of $z\in [0,\; z_T]$, similar procedure can be carried out to cover $z\in [z_T,\; 2z_T]$ and so on \cite{Talbto_Phase_Analytical}.

Figure~\ref{Fig:IdealResults} shows two examples, where the above principal is applied to achieve scaling the spatial period of the input field by a non-integer factor. In the first example, the input field has a period of 25~$\mu$m and the desired increase in the period $m =4/3 = q/p$, so that the output $\Lambda = (4/3)\times 25~\mu$m. The metasurface transfer function $T(x,y)$ is then constructed using \eqref{Eq:PhaseEven} with $n = q = 4$ as shown in the middle of Fig.~\ref{Fig:IdealResults}. The output of the metasurface is then free-space propagated by a distance $\Delta z = (1/p - 1/q)z_T$ leading the output fields, as shown on the right of Fig.~\ref{Fig:IdealResults}, and as expected an increase the period by a factor of $4/3$ is observed. Second example in Fig.~\ref{Fig:IdealResults} shows a similar result showing a period increase by a factor of $3/2$. It should be noted that while these examples only illustrate a period increase by a non-integer factor, similar demonstrations can be easily made for a reduction as well by utilizing fraction Talbot distances between $ [z_T,\; 2z_T]$.

\section{Metasurface Implementation}

Metasurfaces are two dimensional arrays of sub-wavelength electromagnetic scatterers, which are the dimensional reduction of a more general volumetric metamaterial structures. By engineering the electromagnetic properties of the scattering particles, the metasurface can be used to manipulate and engineer the spatial wavefront of the incident waves. By this way, they provide a powerful tool to transform incident fields into specified transmitted and reflected fields \cite{Metasurface_Synthesis_Caloz}. More specifically, metasurfaces can either impart amplitude transformations, phase transformations or both, making them applicable in diverse range of applications involving lensing, imaging \cite{meta2}\cite{meta3}, field transformations \cite{MetaFieldTransformation}, cloaking \cite{MetaCloak} and holograming \cite{MetaHolo}, to name a few. Therefore, considering their versatile field transformation properties and their electrically thin dimensions, they are ideally suited to provide abrupt phase discontinuities in free-space required in the proposed generalized Talbot effect.

\begin{figure}[htbp]
\begin{center}
\subfigure[]{
\psfrag{a}[c][c][0.7]{$\mathbf{p}$}
\psfrag{b}[c][c][0.7]{$\mathbf{m}$}
\psfrag{c}[c][c][0.7]{$2r_1$}
\psfrag{d}[c][c][0.7]{$\Lambda$}
\psfrag{e}[c][c][0.7]{$2r_2$}
\psfrag{f}[c][c][0.7]{$t$}
\psfrag{g}[c][c][0.7]{$n_0$}
\psfrag{h}[c][c][0.7]{$n_h$}
\psfrag{x}[c][c][0.7]{$x$}
\psfrag{y}[c][c][0.7]{$y$}
\psfrag{z}[c][c][0.7]{$z$}
\psfrag{j}[c][c][0.7]{$|R| = 0$}
\includegraphics[width=0.5\columnwidth]{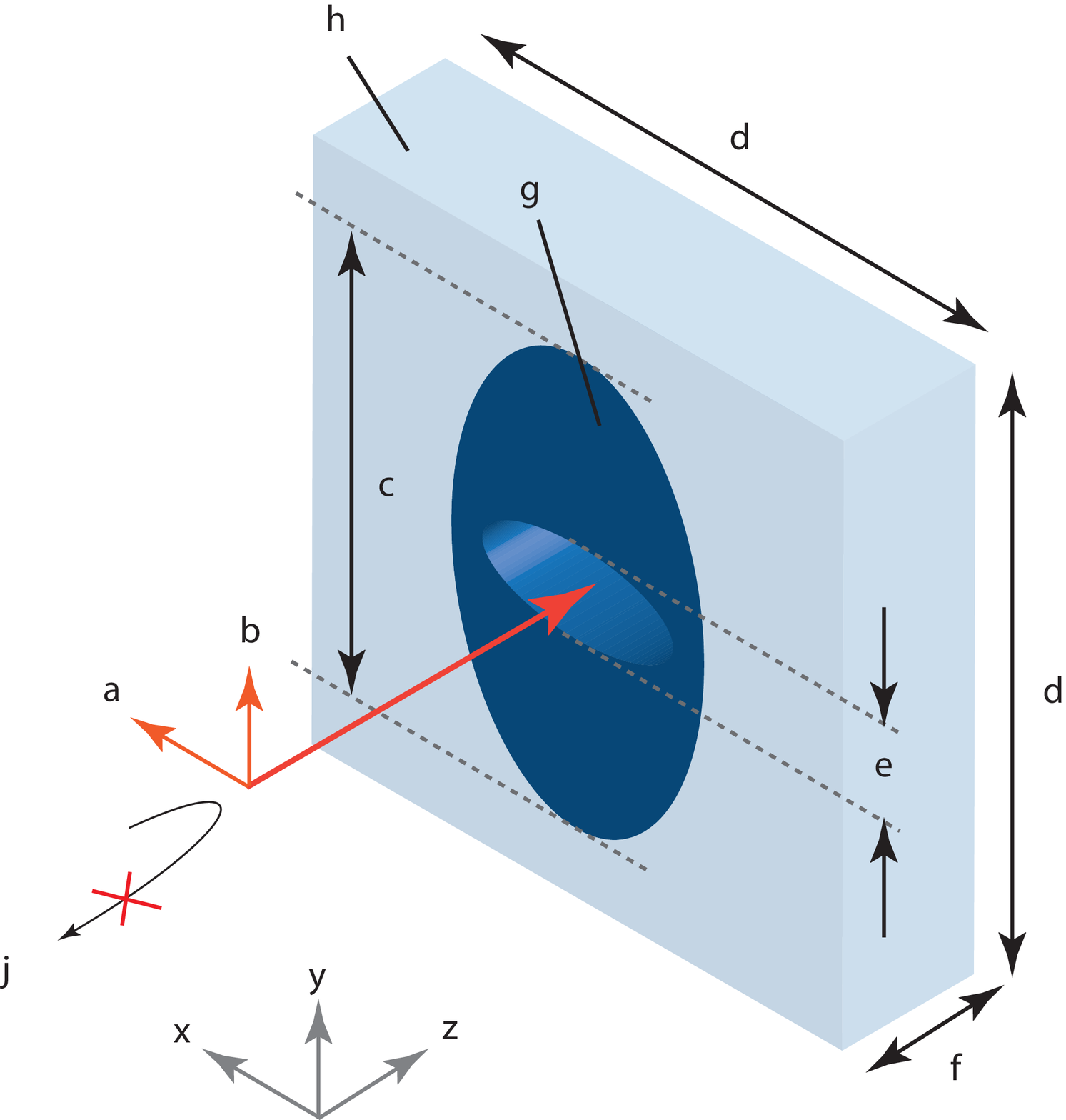}}
\subfigure[]{
\psfrag{a}[c][c][0.8]{frequency $f$ (THz)}
\psfrag{b}[c][c][0.8]{$|T|^2, |R|^2$~(dB)}
\psfrag{d}[c][c][0.8]{$\mathbf{p}$}
\psfrag{c}[c][c][0.8]{$\mathbf{p}$, $\mathbf{m}$}
\psfrag{e}[c][c][0.8]{$\mathbf{m}$}
\psfrag{f}[c][c][0.5]{\shortstack{$r_2=27$~nm\\ $\kappa=0.419$}}
\psfrag{g}[c][c][0.5]{\shortstack{$r_2=55$~nm\\ $\kappa=0.425$}}
\includegraphics[width=0.8\columnwidth]{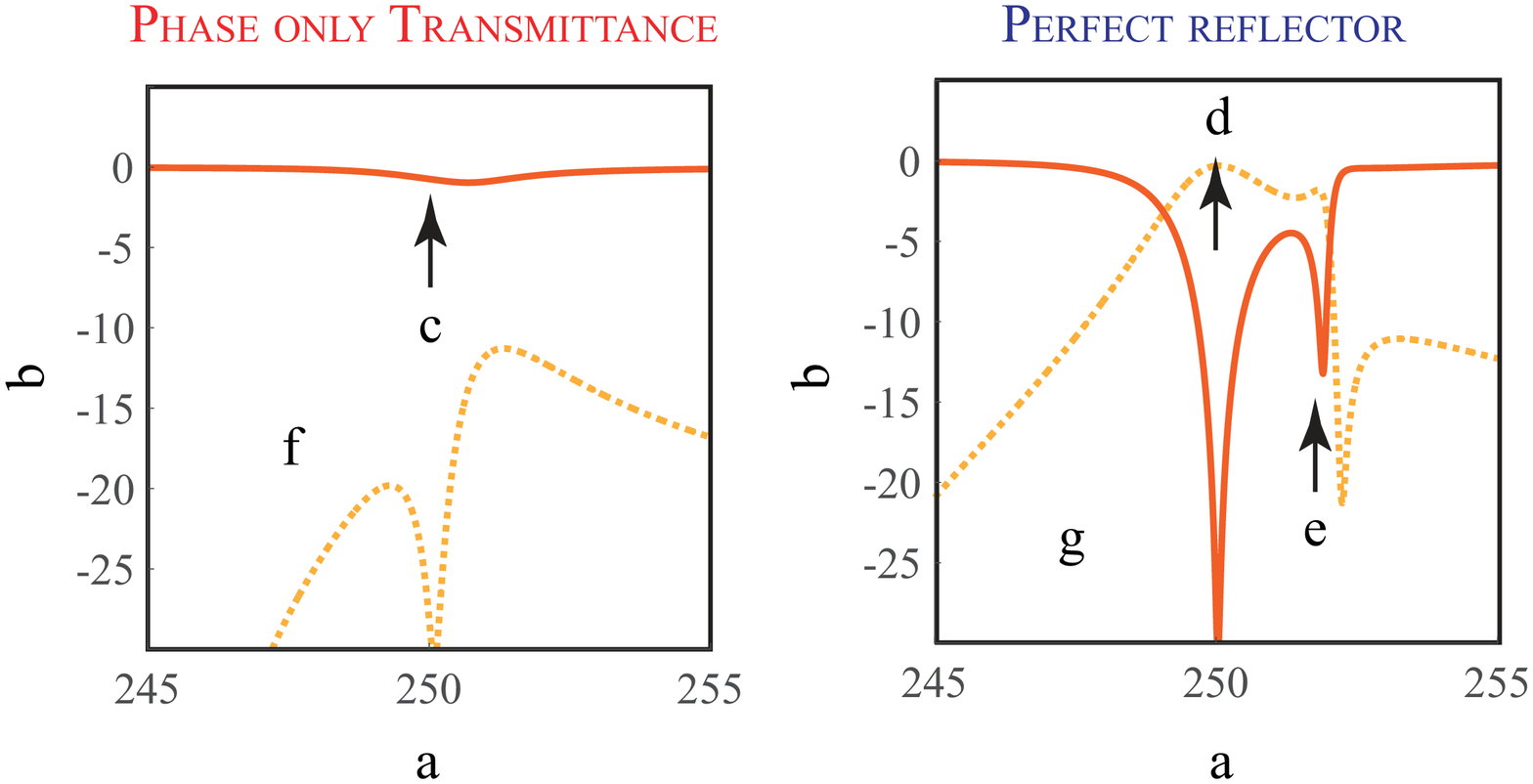}}
\caption{Huygens' source metasurface based on all-dielectric resonators. a) Unit cell periodic in $x-$ and $y-$ directions. b) Typical amplitude transmission (solid) and reflection (dashed) for the two cases when the two dipole moments $\mathbf{p}$ and $\mathbf{m}$ are frequency aligned, and not frequency aligned, respectively. Design parameters: $r_1=300$~nm, $t=220$~nm, $\Lambda = 666$~nm, $n_h= 1.66$~(Silica) and $n_0=3.45$ and $\tan\delta = 0.001$~(Silicon).}\label{Fig:HuyDiMs}
\end{center}
\end{figure}

To provide the needed phase-only filtering characteristics, the metasurface must exhibit ideally a unit amplitude transmission without any reflections, i.e. $|T(x,y)| = 1 \forall \angle T(x,y) \in [0,\;2\pi]$ and reflectance $|R(x,y)| = 0$. These specifications can be conveniently achieved using a so-called \emph{Huygens' metasurface}. A huygens' configurations consists of an orthogonally placed electric and magnetic dipole moments \cite{Kerker_Scattering}, $\mathbf{p}$ and $\mathbf{m}$, respectively, as shown in Fig.~\ref{Fig:HuyDiMs}(a), resulting in a complete cancellation of backscattering as a result of destructive interference of the fields generated by the two dipolar moments. Metasurface consisting of such scattering particles is perfectly matched to free-space and thus has zero reflections, i.e. $|R(x,y)| = 0$. Under lossless conditions, a Huygen's metasurfaces acts as an all-pass surface, with $|T(x,y)| = 1$ and $\angle T(x,y) = \phi_0 \in [0, 2\pi]$. 

\begin{figure*}[htbp]
\begin{center}
\psfrag{X}[c][c][0.75]{$\begin{tabular}{ c|cc } 
				& $r_2$~(nm) & $\kappa$  \\ \hline
 \#1   & 22.5 & 0.4   \\ 
\#2 & 27 & 0.419   \\ 
 \#3& 29 & 0.43   \\
   \hline
\end{tabular}$}
\psfrag{a}[c][c][0.8]{frequency $f$ (THz)}
\psfrag{c}[c][c][0.7]{$|T(x,y)|$~(dB)}
\psfrag{d}[c][c][0.8]{$\angle T(x,y)$~($\pi$~rad)}
\psfrag{f}[c][c][0.8]{$20\log|T|$~(dB)}
\psfrag{e}[c][c][0.7]{Phase $\angle T$ ($\pi$ rad)}
\psfrag{g}[c][c][0.7]{$|\psi(x,y, \Delta z)|^2$~(dB)}
\psfrag{m}[c][c][0.7]{$x~\mu$m}
\psfrag{n}[c][c][0.7]{$y~\mu$m}
\psfrag{h}[c][c][0.7]{$20\log |R| < -10$~dB}
\includegraphics[width=1.7\columnwidth]{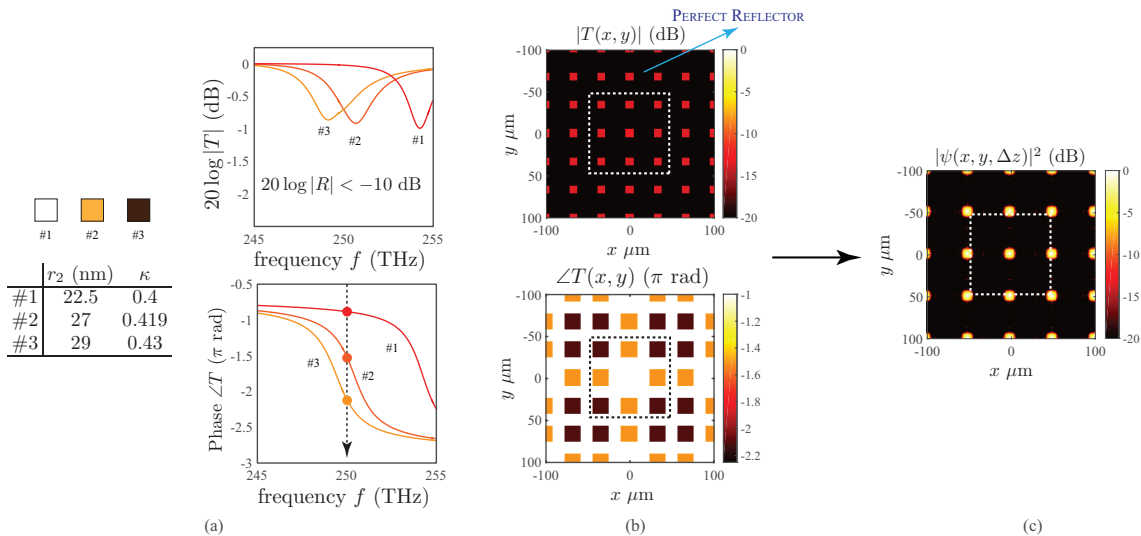}
\caption{Demonstration of the generalized Talbot effect using an all-dielectric metasurface, to achieve  a period scaling by a factor of $m=1.5$, as an example. a) FEM-HFSS simulated transmission and phase responses of three different metasurface unit cells to approximate the required discrete phases. b) Amplitude and phase transmittance of the metasurface aperture using the unit cells of (a). c) The output fields at $\Delta z = (1/p - 1/q)z_T$ under plane-wave excitation of the metasurface aperture computed using Fourier propagation, i.e. $\psi(x,y, \Delta z) = \psi(x,y, 0)\ast h(x,y)$. Only a small part of the overall aperture is shown for clarity. }\label{Fig:HFSS}
\end{center}
\end{figure*}

A practical Huygen's metasurface is conveniently realized using all-dielectric resonator arrays which naturally produce orthogonal $\mathbf{p}$ and $\mathbf{m}$ with lower losses compared to their plasmonic counterparts \cite{AllDielectricKivshar}. A good review on a such all-dielectric metasurfaces can be found in \cite{AllDieelctricMTMS}.  A generalized unit cell of an all-dielectric Huygens' metasurface used in this work is shown in Fig.~\ref{Fig:HuyDiMs}(a), consisting of a high-dielectric holey elliptical puck embedded in a host medium of a lower refractive index $n_h$.
The puck has an ellipticity of $\kappa$ and the hole inside the puck has the elliptical shape as well, but rotated by $90^\circ$. This configuration is particularly useful because its transmission phase at a fixed frequency, can be conveniently tuned by varying the inner radius $r_2$ and $\kappa$ only, without affecting the thickness, lattice size of the unit cell and the outer radius $r_1$, and simultaneously maintaining a good match to free space. 

Fig.~\ref{Fig:HuyDiMs}(b) shows a typical response of such a unit cell for two sets of parameters $r_2$ and $\kappa$, whereby in the first case, the two dipole moments $\mathbf{p}$ and $\mathbf{m}$ are properly excited at the same design frequency ($250$~THz in this example). This results in an optimal interaction of the two dipoles resulting a near-perfect transmission of the wave, as expected from a Huygens' source. This situation corresponds to phase-only transmission response to be used shortly in the generalized Talbot system. The second case, however shows the mis-aligned dipoles resulting in a strong reflection from the unit cell. This situation thus corresponds to a near-perfect reflector.

Using these two configurations, a metasurface aperture can now be constructed to demonstrate a generalized Talbot effect. Let us take an example where the required period scaling at the output plane is $m=1.5 = q/p=3/2$. Since $q=3$ is odd, the discrete phase values are first computed using \eqref{Eq:PhaseOdd}. Next, the metasurface unit cell of Fig.~\ref{Fig:HuyDiMs}(a) is designed to approximate these phase values. Fig.~\ref{Fig:HFSS}(a) shows the transmission and the phase of three such unit cell designs. The reflection in all cases is $< -10$~dB which is sufficiently low in typical practical situations. Using these transmission responses and the perfect reflector unit cell configuration, a metasurface aperture is formed as shown in Fig.~\ref{Fig:HFSS}(b). This completes the metasurface design. A plane wave incidenting on this aperture, and propagating by a distance $\Delta z = (1/p - 1/q)z_T = (1/2 - 1/3)(100~\mu\text{m})^2/\lambda(250~\text{THz})$, transforms into the output fields as shown in Fig.~\ref{Fig:HFSS}(c). As expected and required, the output period is now $50~\mu$m, and thus is $m=1.5$ times more than the one at the input. The metasurface thus successfully performs the specified non-integer period scaling.

\section{Conclusions}
A generalized spatial Talbot effect has been proposed where the period of the input aperture is scaled by a non-integer real value, as opposed to the integer-only factor in a conventional Talbot effect. This has been achieved by engineering phase discontinuity distributions in space using metasurfaces, in conjunction with free-space propagation. Specific implementations using all-dielectric metasurfaces has also been presented and non-integer scalings of the input aperture has been demonstrated using numerical results based on Fourier propagation. While the generalized Talbot effect has been discussed here in the space domain, the proposed principle is equally applicable in the time domain based on space-time duality, where in that case, the repetition rate of the  periodic pulse trains may be scaled by a non-integer factor using temporal phase modulators. 

\bibliography{2016_TalbotDivider_Gupta_PRA}

\end{document}